\documentclass[useAMS,usenatbib,usegraphicx]{mn2e} 
\usepackage{amssymb} 
\usepackage{times} 
\usepackage{aas_macros} 
\usepackage[usenames]{color}

\makeatletter

\newcommand{\Rmnum}[1]{\expandafter\@slowromancap\romannumeral #1@}

\title[Validation of the FM technique]{Validation of the Frequency Modulation technique applied to the pulsating $\delta$~Sct -- $\gamma$~Dor eclipsing binary star KIC~8569819}

\author[Kurtz et al.] 
{
Donald W.~Kurtz$^1$,
Kelly M.~Hambleton$^{1,3}$,
Hiromoto Shibahashi$^2$,
Simon J.~Murphy$^{4,1}$,
\newauthor{Andrej Pr\v sa$^{3,1}$}\\ 
$^{1}$Jeremiah Horrocks Institute, University of Central 
Lancashire, Preston PR1 2HE, UK\\ 
$^{2}$Department of Astronomy, The University of Tokyo, Bunkyo-ku, Tokyo 113-0033, Japan \\
$^{3}$Department of Astrophysics and Planetary Science, Villanova University, Villanova 19087, USA\\
$^{4}$Sydney Institute for Astronomy, School of Physics, Department of Astronomy, The University of Sydney, NSW 2006, Australia\\
}

\begin{document} 

\maketitle 

\begin{abstract} 
KIC~8569819 is an eclipsing binary star with an early F primary and G secondary in a 20.85-d eccentric orbit. The primary is a $\delta$~Sct -- $\gamma$~Dor star pulsating in both p~modes and g~modes. Using 4 years of {\it Kepler} Mission photometric data we independently model the light curve using the traditional technique with the modelling code {\sc phoebe}, and we study the orbital characteristics using the new frequency modulation (FM) technique. We show that both methods provide the equivalent orbital period, eccentricity and argument of periastron, thus illustrating and validating the FM technique. In the amplitude spectrum of the p-mode pulsations we also discovered an FM signal compatible with a third body in the system, a low mass M dwarf in an 861-d orbit around the primary pair. However, the eclipses show no timing variations, indicating that the FM signal is a consequence of the \emph{intrinsic} change in pulsation frequency, thus providing a cautionary tale. Our analysis shows the potential of the FM technique using {\it Kepler} data, and we discuss the prospects to detect planets and brown dwarfs in {\it Kepler} data for A and F stars even in the absence of transits and with no spectroscopic radial velocity curves. This opens the possibility of finding planets orbiting hotter stars that cannot be found by traditional techniques.
\end{abstract} 


\section{Introduction} 
\label{sec:1}

Binary stars are a primary source of fundamental information about stars, particularly their masses and radii. For asteroseismology, modelling of stellar pulsations depends on external determinations of effective temperature and surface gravity, usually from spectroscopy. For heat-driven pulsators where masses and radii cannot be derived from the pulsation frequency spectrum, independent information from eclipsing binary (EB) modelling provides important constraints that narrow the range of possible asteroseismic models. Where pulsating stars are found in binaries, the synergy of the independent techniques from asteroseismology and from the physics of the binary orbit greatly improves our astrophysical inferences about the stars and our confidence in the models that describe them.

The {\it Kepler} Space Mission collected time series light curves of over 190\,000 stars over its four-year main mission lifetime from 2009 to 2013. {\it Kepler} has an orbital period about the Sun of 372.4536~d; during the main mission the satellite performed 4 quarterly rolls (quarters are just over 93~d) per orbit of the Sun. It acquired data of the same field with a $\sim$92\% duty cycle. Its mission is to find extra-solar planets, with emphasis on Earth-like planets and planets in the habitable zone. Its planet candidate list has 4234 entries\footnote{http://kepler.nasa.gov/Mission/discoveries/candidates/} as of August 2014, nearly 1000 of which have been confirmed; in time, 95~per~cent are expected to be confirmed. It also has a list of 2645 eclipsing binary stars\footnote{http://keplerebs.villanova.edu}. More than five hundred main sequence and subgiant solar-like pulsators have been studied asteroseismically with fundamental parameters derived \citep{chaplin2014}. These are critical for the characterisation of extra-solar planets orbiting those stars, so that the asteroseismology and planet studies are synergistic. About 13\,000 red giant stars have been studied asteroseismically (\citealt{stelloeta2013}; \citealt{mosseretal2013}), leading to a better understanding of the stellar structure of giants, and even allowing the determination of core and surface rotation rates \citep{becketal2012}, initiating {\it observational} studies of angular momentum transport in stars with stellar evolution. With {\it Kepler} data, there has now even been an asteroseismic determination of core-to-surface rotation in a main sequence star \citep{kurtzetal14}.

The emphasis of the {\it Kepler} search for habitable planets has been on cool stars, where transits are easier to detect because of the larger planet-to-star size ratio and the proximity of the habitable zones to the host star, hence the shorter orbital periods. The pulsations in cooler stars are stochastically excited by energy in the atmospheric convection zone. The stars show high radial overtone pulsations that are asymptotically nearly equally spaced in frequency, allowing mode identification, and ultimately the extraction of stellar mass, radius and age \citep{aertsetal2010}. The best calibrated cases, such as $\alpha$~Cen (see chapter 7.2.3 of \citealt{aertsetal2010}), allow mass and radius to be determined independently from astrometry and from interferometry. Comparison of the fundamental techniques with the asteroseismic results suggests that asteroseismic masses and radii are as accurate as 2~per~cent in the best cases. 

Hotter stars in {\it Kepler} data are studied less. This is particularly true for the $\delta$~Sct and $\gamma$~Dor stars, where matching models to the observed frequency spectra remains a challenge. Compared with cooler stars, hot stars are not well suited for planetary searches either, for two principal reasons: (i) the transits across the much brighter disks are smaller and more difficult to detect, and can be hidden in the much larger amplitude pulsational variations of most A stars; and (ii) ground-based radial velocity studies to determine the masses of companion exoplanets are more difficult because of the higher masses of the hotter main sequence stars and because of the rotationally broadened spectral lines compared to cooler stars below the Kraft (rotational) break near mid-F spectral type. The first example of a transiting exoplanet orbiting a pulsating $\delta$~Sct star is WASP-33b (HD~15082), where $\delta$~Sct pulsations of amplitude about 1~mmag were found subsequent to the transit discovery \citep{herrero2011}. This star has generated considerable interest with further infrared \citep{deming2012} and optical studies and models (\citealt{kovacs2013}; \citealt{vonessen2014}). The interest is primarily in the use of pulsation characteristics as probes of interactions between the planet and the star. The extrasolar planets encyclopedia\footnote{http://exoplanet.eu} lists only a handful of planets around A stars, the most notable being Fomalhaut~b \citep{kalasetal2013}, the four planets orbiting the $\gamma$~Dor star HR~8799 \citep{marley2012}, and V342~Peg \citep{espositoetal2013}, which have been directly imaged. In consequence, the prevalence of exoplanets orbiting upper main sequence stars is essentially unknown.

\citet{FM2012} developed a new technique for determining orbital parameters of binaries that is based on frequency modulation (FM). This dramatically extends our ability to study binary stars in the {\it Kepler} data set by providing a method that yields traditional `spectroscopic' orbital parameters from photometry alone. Many of the thousands of $\delta$~Sct stars in the {\it Kepler} data set have stable pulsation frequencies. For those stars that are in binary systems, the pulsation frequency is modulated by the orbital motion, producing equally split frequency multiplets in the amplitude spectrum that can be unambiguously identified. \citet{FM2012} show how these multiplets can be used to determine the orbital frequency, the mass function (as in a spectroscopic single-lined binary star), $a \sin i$ for the pulsating primary star, and the eccentricity. 
More recently, Shibahashi, Kurtz and Murphy (in preparation) have extended the technique to include the determination of the argument of periastron. These are all parameters that in the past required a large spectroscopic data set to determine radial velocities. Recently, \citet{murphyetal14} developed an analogous technique based on phase modulation (PM). This technique is equivalent to FM, and is more easily automated.

In this paper we focus on KIC~8569819, an eclipsing binary in an eccentric orbit with a primary star that is a pulsating $\delta$~Sct -- $\gamma$~Dor star. We derive orbital parameters independently from both the eclipsing binary light curve fitting, and from the FM technique.

\begin{figure*}
\centering	
\includegraphics[width=1.0\linewidth,angle=0]{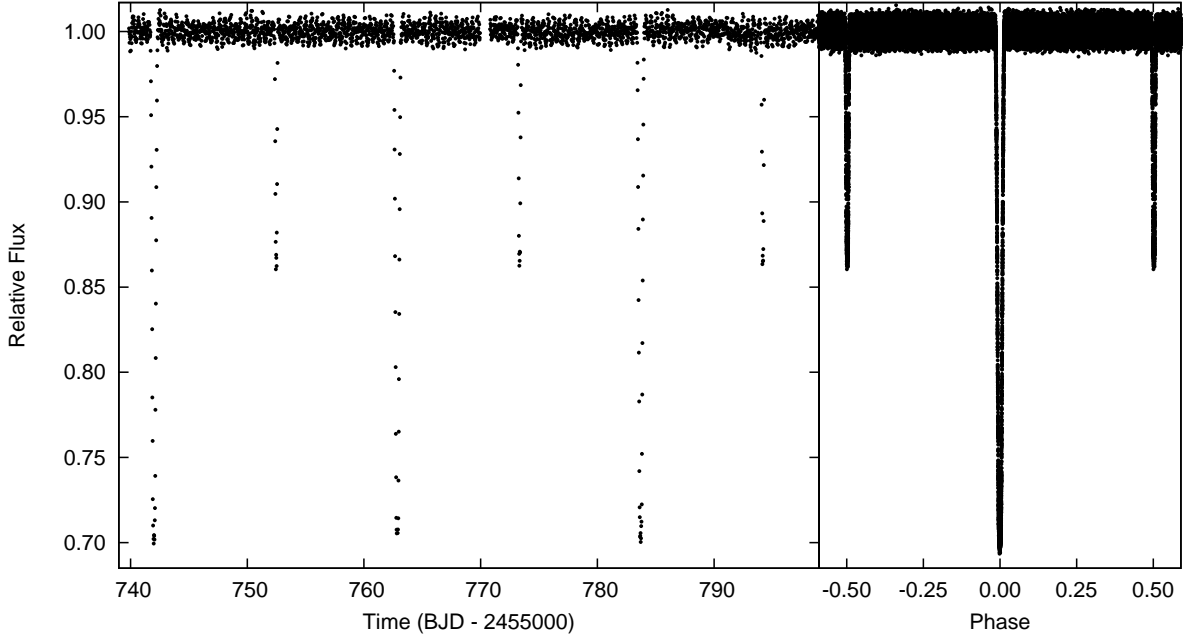}
\caption{A phased light curve of KIC~8569819 for the Q0--17 msMAP LC data with no further processing. The left panel shows three cycles of the 20.84-d binary orbit with primary and secondary eclipses. The right panel shows the full Q0--17 LC data set phased on the orbital period. }
\label{fig:8569819_lc}
\end{figure*}

\section{KIC~8569819: an eclipsing binary FM star}
\label{sec:8569819}

KIC~8569819 is a $Kp = 13.0$ eclipsing binary with $T_{\rm  eff} = 7100$~K and  $\log g = 4.0$ (cgs units) in the {\it Kepler} Input Catalogue (KIC) ({see \citet{huber2014} for a discussion of errors in the KIC; at this temperature and surface gravity they are about $\pm 250$\,K in $T_{\rm eff}$ and $\pm 0.2$ in $\log g$).} The contamination parameter is 0.237, but a visual check of the pixel-level data shows that the mask used in the reductions does not include the nearest possible contaminating star. 

The data used for the analysis in this paper are {\it Kepler} quarters 0 to 17 (Q0--Q17) long cadence (LC) data with 29.4-min integration times. We used the multi-scale, maximum a posteriori (msMAP) pipeline data; information on the reduction pipeline can be found in the {\it Kepler} data release notes\footnote{https://archive.stsci.edu/kepler/data$\_$release.html} 21. Fig.~\ref{fig:8569819_lc} depicts a light curve for KIC~8569819 for a section of the LC msMAP data where we can see both primary and secondary eclipses. The separation between the eclipses is close to 0.5, but closer examination shows that primary eclipse lasts for about 13~h, and secondary for 6.5~h, requiring a high eccentricity of $e \approx 0.4$, detailed in Section~\ref{sec:ebmodel}. The eclipses are flat-bottomed (total), therefore $i \approx 90^{\circ}$. The primary eclipse takes longer: it occurs near apastron, with the cooler companion being in front. 

The FM analysis of KIC~8569819 is presented in Section~\ref{sec:fm} and the eclipsing binary light curve analysis in Section~\ref{sec:ebmodel}. These two analyses were performed independently for objective comparison of the results. 

\section{Frequency modulation analysis of KIC~8569819}
\label{sec:fm}

\subsection{Frequency modulation of $\nu_1$: the 20.85-d binary}

For the analysis of the pulsation frequencies we have masked out the eclipses from the light curve. This is necessary because of the high amplitude peaks they generate at low frequency in the amplitude spectrum; these have spectral window patterns that extend out to the $\delta$~Sct range of the pulsation frequencies. The pulsation amplitudes also change during eclipses because of the changing background light level, and because of the partial obscuration of the pulsating star during ingress and egress of the primary eclipse. That generates amplitude modulation sidelobes to the pulsation peaks separated by exactly the orbital frequency, hence overlapping with the FM signal that we are studying. Unless the binary star model encompasses a full description of the pulsations, masking the data-set is preferred to subtracting a binary model fit alone.

\begin{figure}
\centering
\includegraphics[width=0.9\linewidth,angle=0]{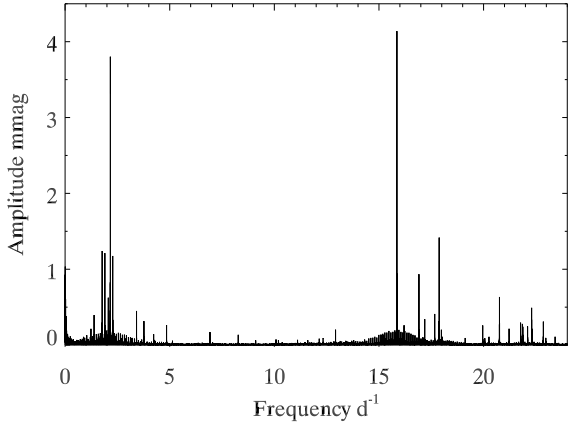}
\caption{Amplitude spectrum for the masked Q0--Q17 light curve. In the low-frequency range between $0-5$~d$^{-1}$ there are g-mode pulsation frequencies. In the high-frequency range between $12-24$~d$^{-1}$ there are p-mode peaks.}
\label{fig:8569819_ft-all}
\end{figure}

Fig.~\ref{fig:8569819_ft-all} shows the amplitude spectrum out to nearly the Nyquist frequency ($\sim$24.5~d$^{-1}$) for KIC~8569819 for the masked Q0--17 LC data. There are pulsations in both the g-mode and p-mode frequency regions. Since the relative amplitude of the FM sidelobes to the amplitude of the central peak, i.e., the detectability of the FM signal, is proportional to $\nu$, we concentrate our analysis only on the p-mode frequency range. The low, equally-spaced combs of peaks around the highest pulsation peaks in Fig.~\ref{fig:8569819_ft-all} are part of the window function resulting from the masking of the light curve. Those are spectral window sidelobes at the orbital frequency, but they are removed by prewhitening of the main peak and leave no trace in the amplitude spectrum of the residuals, hence they do not perturb our analysis.

\begin{figure}
\centering
\includegraphics[width=0.9\linewidth,angle=0]{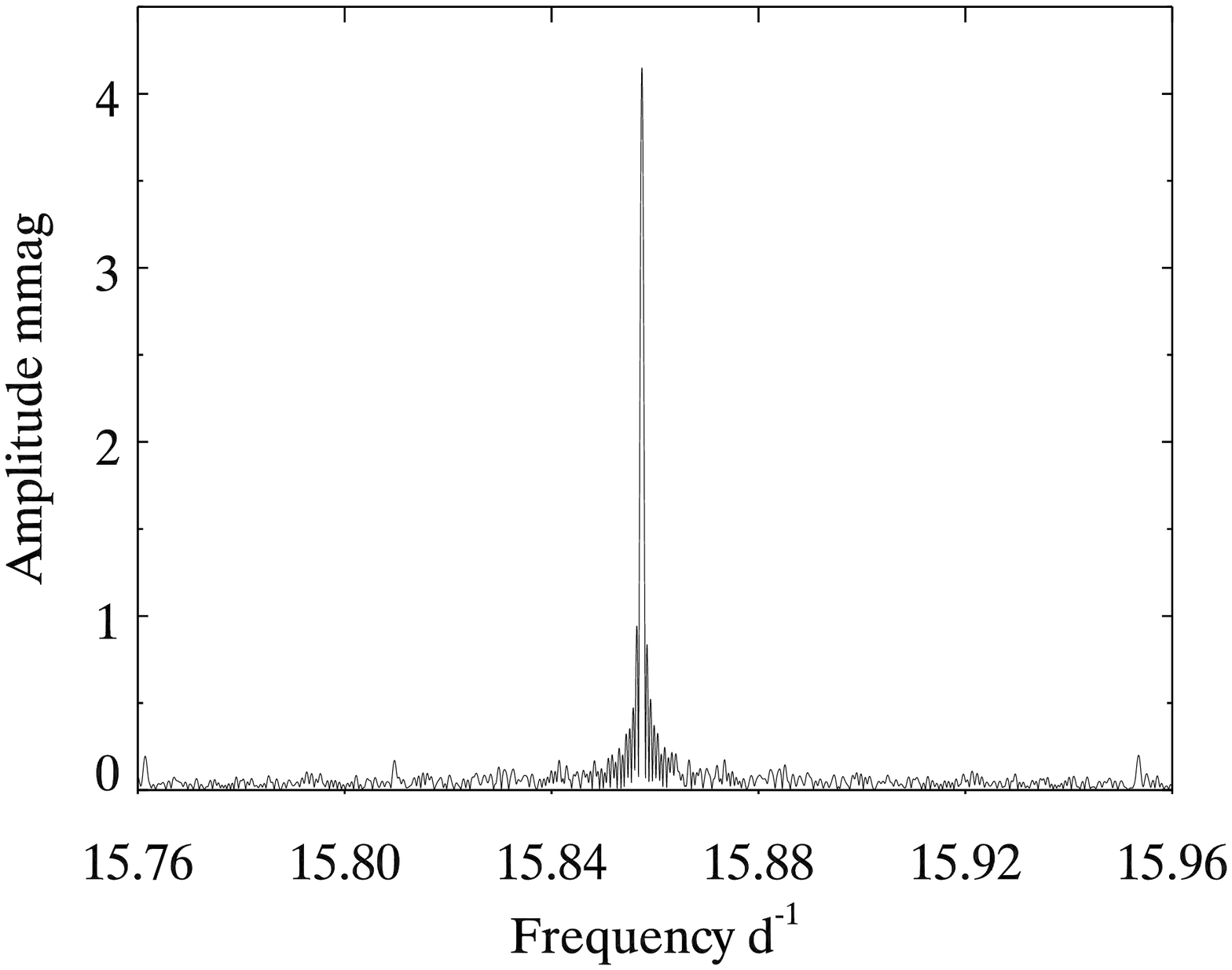}
\includegraphics[width=0.9\linewidth,angle=0]{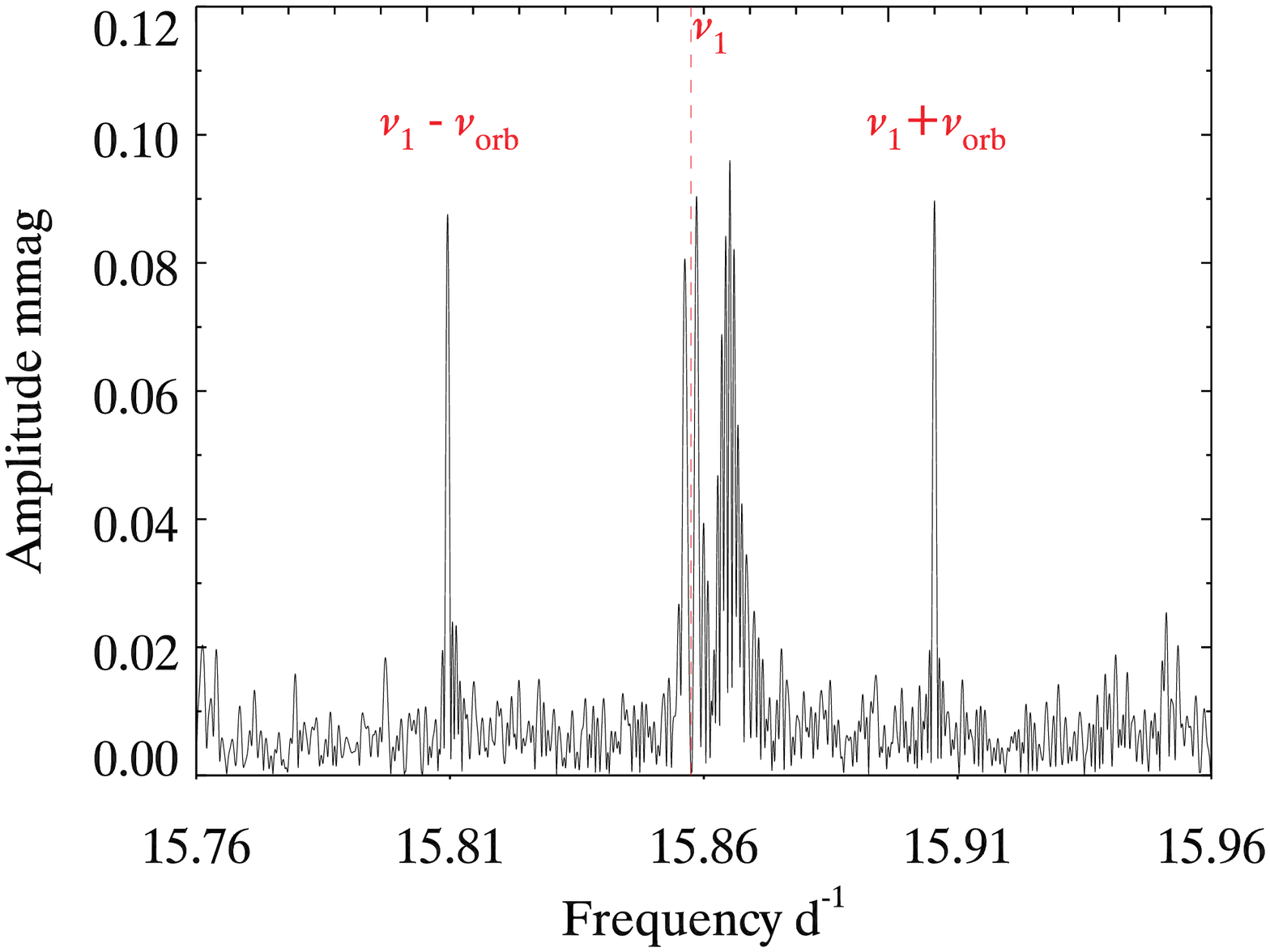}
\caption{Top: An amplitude spectrum for the masked Q0--Q17 light curve in the $\delta$~Sct frequency range of the highest amplitude p~mode. Bottom: After prewhitening $\nu_1$. There are important peaks on either side of $\nu_1$ that are discussed in the text. Here, note the equally spaced sidelobes near $\nu = 15.81$~d$^{-1}$ and 15.91~d$^{-1}$ that are separated from $\nu_1$ by $\nu_{\rm orb}$. These are the FM first sidelobes. There is some visual indication of the presence of the second sidelobes.}
\label{fig:8569819_ft}
\end{figure}

The highest amplitude $\delta$~Sct p~mode is at a frequency $\nu_1 = 15.8574687(5)$~d$^{-1}$ and is shown in the top panel of Fig.~\ref{fig:8569819_ft}. For an estimate of the p-mode radial overtone, it is useful to calculate the $Q$ value for $\nu_1$. This is defined to be:
\begin{equation}
	Q = {P_{\rm osc}}\sqrt{\frac{\overline{\rho}}{{\overline{\rho}_{\odot}}}}
\label{eq:4}
\end{equation}
where $P_{\rm osc}$ is the pulsation period and $\overline\rho$ is the mean density. $Q$ is known as the `pulsation constant'. 
{Using the definition of mean density as $\overline\rho = \frac{M}{\frac{4}{3}\upi R^3}$, surface gravity as $g = \frac{GM}{R^2}$, absolute luminosity as $L = 4 \upi R^2 \sigma T_{\rm eff}^4$, and absolute magnitude as $M_{\rm bol} = -2.5 \log L + {\rm constant}$,  
}
equation~(\ref{eq:4}) can be rewritten as:
\begin{equation}
\log Q = -6.454 +\log P_{\rm osc} +\frac{1}{2}\log g +\frac{1}{10}M_{\rm bol} + \log T_{\rm eff}, 
\label{eq:30}
\end{equation}
where $P_{\rm osc}$ is given in d, $\log g$ is in cgs units and $T_{\rm  eff}$ is in K. Using the KIC values of $T_{\rm  eff} = 7100$~K and $\log g = 4.0$, and estimating the bolometric magnitude to be 2.8, we obtain $Q =0.030$, typical of fundamental to first overtone pulsation in $\delta$~Sct stars \citep{stellingwerf1979}. We thus conclude that the p-mode frequencies are likely to be due to low overtone modes.  Prewhitening the data by $\nu_1$ gives the amplitude spectrum of the residuals depicted in the bottom panel of Fig.~\ref{fig:8569819_ft}, where the first FM orbital sidelobes are annotated. 

Table~\ref{table:lsfit_8569819} shows the nonlinear least-squares fit of $\nu_1$ and its first FM orbital sidelobes. The formal errors from the least-squares fit include all of the variance in the data, which (as can be seen in Fig.~\ref{fig:8569819_ft-all}) includes astrophysical variance due to all pulsation frequencies in both the g-mode and p-mode regions. To estimate the intrinsic scatter, we focused on a featureless section of the amplitude spectrum in the range 9--10~d$^{-1}$. The highest peaks in that region have amplitudes of $\sim$20~$\mu$mag. For an amplitude spectrum with this density of frequencies, even for normally distributed data we expect numerous peaks with amplitudes greater than 3$\sigma$, and in practice we find that the highest amplitude peaks have amplitudes about four times the formal amplitude error for the least-squares fit. We therefore estimate the amplitude error to be 0.005~mmag (20~$\mu$mag/4). Since phase and frequency errors scale with amplitude error \citep{montgomery1999}, we scale the formal errors for frequency and phase in Table~\ref{table:lsfit_8569819} by the same factor as the revised error in amplitude.

The frequency triplet in Table~\ref{table:lsfit_8569819} is equally split; the two splittings agree to 0.6$\sigma$, suggesting that our reduced errors are conservative. The average of the two splittings is the orbital frequency, $\nu_{\rm orb} = 0.047964  \pm 0.000014$~d$^{-1}$. This differs by 0.14$\sigma$ from $\nu_{\rm orb} = 0.047962$~d$^{-1}$ obtained by phase-folding the eclipses (cf.~Fig.~\ref{fig:8569819_lc}). Phase-folding provides a more precise orbital frequency than the FM signal because of the much higher signal-to-noise ratio for the eclipses compared to the orbital pulsational frequency shifts. The agreement between the two methods is excellent. 

Next we re-fit the frequency multiplet by forcing the splitting to be exactly equal. There is no significant difference to the result whether the orbital period is chosen from the frequency splitting or from phase-folding, so to keep the FM analysis independent, we use the value from Table~\ref{table:lsfit_8569819}. For the study of orbital characteristics using the FM technique \citep{FM2012} it is important that the splittings are exactly equal; we are then able to use the phase information, since frequency and phase are coupled in the Fourier sinusoidal description. 

This coupling between frequency and phase is easy to see. The function we fit to our data has the form $\cos (\omega t + \phi)$, where $\omega = 2 \upi \nu$ is the angular frequency. If we imagine a change to the angular frequency such that $\omega^\prime = \omega + \delta\omega$, then we can write the argument of the cosine function to be $\omega^\prime t + \phi = (\omega + \delta\omega) t + \phi = \omega t + (\phi + \delta\omega t) = \omega t + (\phi + \delta\phi) = \omega t + \phi^\prime$, where $\phi^\prime = \phi + \delta\omega t$. It is not possible to distinguish between a change $\delta\omega$ to the frequency or a change $\delta\phi$ to the phase without an external constraint.  

\begin{table*}
\centering
\caption[]{A nonlinear least-squares fit of the highest amplitude frequency seen in Fig.~\ref{fig:8569819_ft} and its first orbital FM sidelobes to the Q0--Q17 LC masked {\it Kepler} data for KIC~8569819. The frequencies are separated by $\nu_{\rm orb} = 0.047964  \pm 0.000014$~d$^{-1}$. The zero point in time for the phase is BJD~2455672.2. }
\begin{tabular}{lr@{~$\pm$~}lcr}
\hline
&\multicolumn{2}{c}{frequency} & \multicolumn{1}{c}{amplitude} &
\multicolumn{1}{c}{phase} \\
&\multicolumn{2}{c}{d$^{-1}$} & \multicolumn{1}{c}{mmag} &
\multicolumn{1}{c}{radians}\\
\hline
$\nu_1 - \nu_{\rm orb}$  & $15.8095167$ &  $0.0000211  $ & $  0.089  \pm 0.005  $ & $  2.9067  \pm 0.0561  $   \\  
$\nu_1$  & $15.8574721$ & $0.0000005  $ & $  4.150  \pm 0.005  $ & $  2.4425  \pm 0.0012  $ \\  
$\nu_1 + \nu_{\rm orb}$  & $15.9054555$  & $0.0000197  $ & $  0.096  \pm 0.005  $ & $  -1.2593  \pm 0.0524  $ \\
\\ 
$\nu_{\rm orb_1} = \nu_1 - (\nu_1 - \nu_{\rm orb})$  & $  0.047955$ &  $0.000021$  & &\\
$\nu_{\rm orb_2} = (\nu_1 + \nu_{\rm orb}) - \nu_1$  & $  0.047973$ &  $0.000020$  & &\\
$\nu_{\rm orb_1} -\nu_{\rm orb_2}$  & $  0.000018$ & $0.000029$   & &\\  
$\nu_{\rm orb} \equiv \langle \nu_{\rm orb_1},\nu_{\rm orb_2} \rangle$ &  $  0.047964$ &  $0.000014$  & & \\  
$P_{\rm orb}$ (d)&  $  20.849$ & $0.006$   & &\\  
\hline
\end{tabular}
\label{table:lsfit_8569819}
\end{table*}

Table~\ref{table:lsfit2_8569819} lists the results of fitting the equally-spaced quintuplet for the highest amplitude $\delta$~Sct mode to the data. The zero point in time, ${\rm BJD}2455679.12090 \pm 0.26$, has been chosen to set the phases of the first FM sidelobes equal; the error in the zero point is derived from 1$\sigma$ in the difference between the sidelobe phases. This zero-point corresponds to the time when the motion of the stars is perpendicular to the line of sight. Thus, by measuring the difference between the zeropoint time and the time of superior conjunction, we can derive the argument of periastron. The time difference can be converted to a phase difference, and the phase difference, via mean and eccentric anomaly, converted to true anomaly. The true anomaly is a measure of the angle between the point of the orbit where $v_r = 0$ and superior conjunction; a simple Keplerian integrator can be employed to solve this for the argument of periastron, which yields $\omega = 4.59 \pm 0.16$~rad. 

The difference between the first sidelobe phases and the central peak phase is $-1.62 \pm 0.06$~rad, which is equal to $-\upi/2$ within the errors, as expected from the FM theory. The minus sign indicates that the line-of-sight crossing of the pulsating star occurs on the far side of the orbit for the chosen time zero-point, $t_0$. This phase relationship demonstrates that the frequency triplet describes pure frequency modulation, as expected for FM.

\citet{FM2012} define a parameter $\alpha$ that measures the amplitude of phase modulation when the pulsation frequency is treated as fixed (Eq.~(4) therein). When $\alpha \ll 1$, it is given by 

\begin{equation} \label{eq:1}
\alpha = \frac{A_{+1}+A_{-1}}{A_0},
\end{equation}

\noindent where $A_{+1}$ and $A_{-1}$ are the observed amplitudes of the first FM sidelobes, and $A_0$ is the observed amplitude of the central peak of the FM multiplet. From Table~\ref{table:lsfit2_8569819} we then find for KIC~8569819 that $\alpha = 0.0444 \pm0.0017$. From this amplitude ratio, the pulsation period and the orbital period, the mass function can be derived:

\begin{equation} \label{eq:2}
f(m_1,m_2,\sin i) \equiv {{(m_2 \sin i)^3}\over{(m_1 + m_2)^2}} = \alpha^3  \frac{P_{\rm osc}^{3}}{P_{\rm orb}^{2}} \frac{c^3}{2 \upi G}.
\end{equation}

\noindent Using the data in Table~\ref{table:lsfit2_8569819}, we find $f(m_1,m_2,\sin i) = 0.141 \pm 0.016$~M$_{\odot}$. Adopting a typical mass of $m_1 = 1.7$~M$_{\odot}$ for the primary and $i \approx 90^{\circ}$, a secondary mass is $m_2 = 1$~M$_{\odot}$, hence the companion to the $\delta$~Sct star is probably a solar-like main sequence star. 

We can also derive the semi-major axis of the primary star about the barycentre. That is given by:
\begin{equation} \label{eq:3}
a_1\sin i ={{P_{\rm osc}}\over{2\upi}} \alpha c,
\end{equation}
\noindent from which we find $a_1 \sin i = 0.0772 \pm 0.0030$~au. 

\begin{table*}
\centering
\caption[]{A least-squares fit of the frequency quintuplet for the highest amplitude mode to the Q0--Q17 LC {\it Kepler} data for KIC~8569819. The frequencies of the multiplet are separated by the orbital frequency, $\nu_{\rm orb} = 0.047964  \pm 0.000014$~d$^{-1}$ ($P_{\rm orb} = 20.849  \pm 0.006$~d). The zero point for the phases has been chosen to be a time when the phases of the first sidelobes to the highest amplitude frequency are equal,  $t_0 = {\rm BJD}~2455679.12090$. It can be seen that the phases of the first sidelobes differ from that of the phase of $\nu_1$ by $-1.62 \pm 0.06$~rad, which is equal to $\upi/2$ as required by the theory \citep{FM2012}. The mass function and orbital eccentricity are derived from the sidelobes' amplitudes. }
\begin{tabular}{lccr}
\hline
&\multicolumn{1}{c}{frequency} & \multicolumn{1}{c}{amplitude} &
\multicolumn{1}{c}{phase} \\
&\multicolumn{1}{c}{d$^{-1}$} & \multicolumn{1}{c}{mmag} &
\multicolumn{1}{c}{radians}\\
\hline
$\nu_1 - 2\nu_{\rm orb}$ & 15.7615434  &  $0.014 \pm 0.005 $ & $-3.0812 \pm 0.2775$  \\ 
$\nu_1 - \nu_{\rm orb}$ & 15.8095077  &  $0.089 \pm 0.005 $ & $-0.7630 \pm 0.0561 $ \\ 
$\nu_1$ & 15.8574721  &  $4.148 \pm 0.005 $ & $0.8593 \pm 0.0012 $ \\ 
$\nu_1 + \nu_{\rm orb}$ & 15.9054365 &  $0.096 \pm 0.005 $ & $-0.7630 \pm 0.0541$  \\ 
$\nu_1 + 2 \nu_{\rm orb}$  & 15.9534009  &  $0.022 \pm 0.005 $ & $2.9350 \pm 0.2029 $ \\ 
\hline
\end{tabular}
\label{table:lsfit2_8569819}
\end{table*}

To derive the eccentricity of the system, we have fitted a frequency quintuplet split by the orbital frequency about $\nu_1$ to the Q0--17 LC data as shown in Table~\ref{table:lsfit2_8569819}. While the second FM sidelobes in Fig.~\ref{fig:8569819_ft} are only marginally visible, the least-squares fit shows them to be significant to 2.8$\sigma$ and 4.4$\sigma$. The False Alarm probability of finding peaks of this significance at {\it particular} frequencies (i.e., the second FM sidelobes) is $F = \exp(-z)$ \citep{hornebaliunas1986}, where $z$ is the power signal-to-noise ratio. Taking an average amplitude signal-to-noise ratio of 3.6 for our second FM sidelobes gives $z = 12.96$ and $F = 2.4 \times 10^{-6}$. Hence we can use the FM sidelobes to derive the eccentricity:
\begin{equation} \label{eq:6}
	e =  {{2(A_{+2}+A_{-2})}\over{(A_{+1}+A_{-1})}} = 0.39 \pm 0.08,
\end{equation}
\noindent where $A_{+2}$ and $A_{-2}$ are the observed amplitudes of the second FM sidelobes. While the phases of the second sidelobes contain information about the argument of periastron, the errors are too large to use them in this case.

\subsection{Further frequency modulation of $\nu_1$: a cautionary tale}
\label{sec:cautionarynote}

\begin{figure}
\centering
\includegraphics[width=0.9\linewidth,angle=0]
{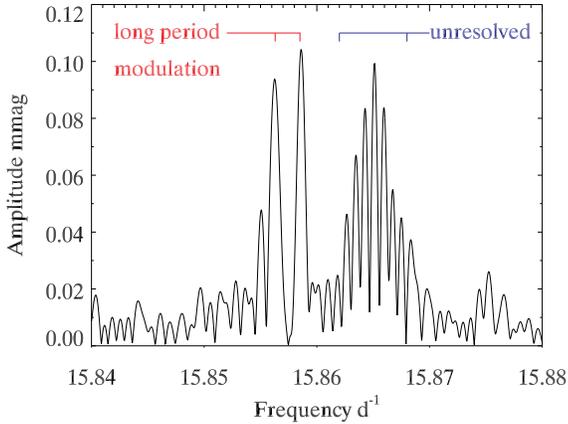}
\caption{A zoomed-in region around $\nu_1$ after prewhitening. The multiplet of peaks to the right is most likely the result of a low, amplitude-independent pulsation mode that is modulated on a time scale longer than the data span, since the sidelobes are not fully resolved. The other two central peaks are equally spaced about $\nu_1$ and are caused by pure frequency modulation with a period of 861-d.}
\label{fig:8569819_ft2}
\end{figure}

We now return to the peaks in the immediate vicinity of $\nu_1$ (cf.~bottom panel of Fig.~\ref{fig:8569819_ft}, and the zoomed prewhitened region in Fig.~\ref{fig:8569819_ft2}). A multiplet of peaks and a doublet can be seen in the amplitude spectrum after $\nu_1$ has been prewhitened. The multiplet on the right consists of several unresolved peaks that are likely caused by a low amplitude ($\sim$10~$\mu$mag) independent pulsation mode that is amplitude-modulated on a time scale longer than the 4-yr data set. This kind of amplitude modulation is commonly seen for $\delta$~Sct stars in the {\it Kepler} data (see, e.g., \citealt{bowman2014}), hence we discuss these frequencies no further here. 

On the other hand, the doublet seen in Fig.~\ref{fig:8569819_ft2} is fully resolved from $\nu_1$ and equally spaced on either side. By repeating the phase relationship exercise done on the outer set of sidelobes, we find that the peaks are caused by pure frequency modulation with a modulation period of 861~d. It is tempting to conclude that these peaks are the FM sidelobes caused by a third companion orbiting the binary. Tables~\ref{table:lsfit_8569819_md} and \ref{table:lsfit2_8569819_md} show the fit of the inner sidelobes to $\nu_1$. The modulation period is derived to be $861 \pm 11$~d. If we assumed that this frequency modulation is caused by a third body, then from the amplitudes of the orbital sidelobes given in Table~\ref{table:lsfit2_8569819_md} it would follow that $\alpha = 0.0441 \pm 0.0017$, from eq.~5 $a \sin i = 38$~light~seconds, and from eq.~4 the mass function is $f(m_1+m_2,m_3,\sin i) = 0.000119 \pm 0.000012$~M$_{\odot}$. Using the derived mass of the binary, $m_1 + m_2 = 2.7$~M$_{\odot}$ and $i \sim 90^{\circ}$, we would obtain a tertiary mass of $m_3 = 0.098 \pm 0.003$~M$_{\odot}$, i.e.~a low mass main sequence M dwarf star. We examine this proposition now, and show it to be incorrect. 

\begin{table*}
\centering
\caption[]{A nonlinear least-squares fit of the highest amplitude frequency seen in Fig.~\ref{fig:8569819_ft} and its first closely-spaced frequency modulation sidelobes seen in Fig.~\ref{fig:8569819_ft2} to the Q0--Q17 LC masked {\it Kepler} data for KIC~8569819. The frequencies are separated by $\nu_{\rm mod} = 0.001161  \pm 0.000015$~d$^{-1}$, giving an modulation period of $861 \pm 11$~d. The zero point in time for the phase is BJD~2455679.12920.}
\begin{tabular}{lr@{~$\pm$~}lcr}
\hline
&\multicolumn{2}{c}{frequency} & \multicolumn{1}{c}{amplitude} &
\multicolumn{1}{c}{phase} \\
&\multicolumn{2}{c}{d$^{-1}$} & \multicolumn{1}{c}{mmag} &
\multicolumn{1}{c}{radians}\\
\hline
$\nu_1 - \nu_{\rm mod}$ & $15.8562606$ & $0.0000216$ & $0.089 \pm 0.005$ & $ -1.5634 \pm 0.0570$ \\ 
$\nu_1$ & $15.8574741$ & $0.0000005$ & $4.152 \pm 0.005$ & $0.0284 \pm 0.0016$ \\ 
$\nu_1 + \nu_{\rm mod}$  & $15.8585831$ & $0.0000204$ & $0.096 \pm 0.005$ & $-1.2470 \pm 0.0529$ \\ 
\\ 
$\nu_{\rm mod_1}$  & $  0.001214$ & $0.000022$  & &\\
$\nu_{\rm mod_2}$  & $  0.001109$ & $0.000020$  & &\\
$\nu_{\rm mod_1} -\nu_{\rm mod_2}$  & $  0.00006$ & $0.00003$   & &\\  
$\nu_{\rm mod} \equiv \langle \nu_{\rm mod_1},\nu_{\rm mod_2} \rangle$ &  $  0.001161$ & $0.000015$  & & \\  
$P_{\rm mod}$ (d)&  $  861$ & $11$   & &\\  
\hline
\end{tabular}
\label{table:lsfit_8569819_md}
\end{table*}

\begin{table*}
\centering
\caption[]{A linear least-squares fit of the close frequency triplet for the highest amplitude mode to the Q0--Q17 LC {\it Kepler} data for KIC~8569819 with exactly equal splitting. The frequencies of the triplet are separated by the modulation frequency, $\nu_{\rm mod} = 0.001157  \pm 0.000014$~d$^{-1}$ ($P_{\rm mod} = 861  \pm 11$~d). The zero point for the phases has been chosen to be a time when the phases of the first sidelobes to the highest amplitude frequency are equal,  $t_0 = {\rm BJD}~2455347.56763$. It can be seen that the phases of the first sidelobes differ from that of the phase of $\nu_1$ by $-1.56 \pm 0.03$~rad, which is equal to $-\upi/2$, proving pure frequency modulation. }
\begin{tabular}{lccr}
\hline
&\multicolumn{1}{c}{frequency} & \multicolumn{1}{c}{amplitude} &
\multicolumn{1}{c}{phase} \\
&\multicolumn{1}{c}{d$^{-1}$} & \multicolumn{1}{c}{mmag} &
\multicolumn{1}{c}{radians}\\
\hline
$\nu_1 - \nu_{\rm mod}$ & $15.856313$ & $0.088 \pm 0.005$ & $1.8345 \pm 0.0560$ \\ 
$\nu_1$ & $15.857474$ & $4.149 \pm 0.005$ & $-2.8897 \pm 0.0012$ \\ 
$\nu_1 + \nu_{\rm mod}$ & $15.858635$ & $0.095 \pm 0.005$ & $1.8344 \pm 0.0522$ \\ 
\hline
\end{tabular}
\label{table:lsfit2_8569819_md}
\end{table*}

Many $\delta$~Sct stars show pulsation modes that are amplitude-modulated, but modes can also be intrinsically \emph{frequency}-modulated (e.g., \citealt{bowman2014}; \citealt{breger00}). Thus, another interpretation of the close frequency sidelobes to $\nu_1$ is that they represent \emph{intrinsic} (i.e.~non-dynamical) frequency modulation of that pulsation mode on a timescale of 861~d. A test to discriminate between intrinsic and dynamical FM is to look for the sidelobes in \emph{several} pulsation frequencies. In the case of dynamical FM (3rd body), all pulsation frequencies must show the same FM signature, akin to that in Fig.~\ref{fig:8569819_ft}. In the case of intrinsic FM, different frequencies will have different sidelobes, corresponding to different pulsation cavities in the star. 

\begin{figure}
\centering
\includegraphics[width=\linewidth]{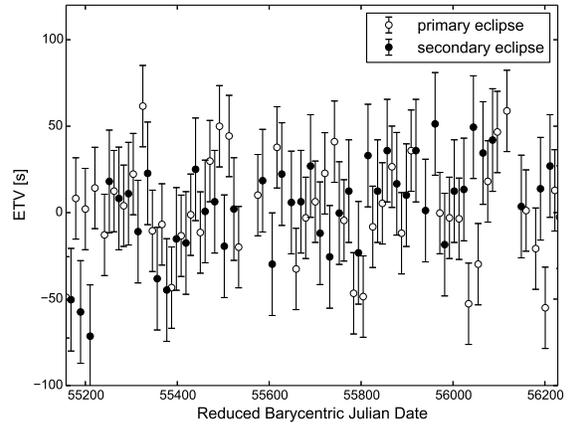} \\
\caption{Eclipse timing variations for KIC~8569819. Eclipse times are measured by fitting a polynomial chain to the entire phased light curve, and using that function to fit the time offset of each successive eclipse \citep{conroy2014}. Primary ETVs are depicted with open circles and secondary ETVs are depicted with filled circles.}
\label{fig:etv}
\end{figure}

Unfortunately, all other p~mode amplitudes are at least a factor of 4 or more smaller than the amplitude of $\nu_1$ (cf.~Fig.~\ref{fig:8569819_ft-all}), hence we do not have sufficient signal in other frequencies to test for the very low amplitudes of closely spaced FM sidelobes expected for an 861-d orbital period. While asteroseismology might not provide a definitive answer, we do have another test: eclipse timing variations (ETVs; \citealt{conroy2014}). When a binary star is in a gravitationally bound system with another body, its center of mass will move around the system's barycenter. Because of the light time travel effect, the time of eclipses depends on the binary star's position on the outer orbit. If a third body were present in the system, then the eclipsing binary would be separated by $a \sin i = 38$~light seconds from the barycenter, hence we would see a clear signal of that amplitude in eclipse timings. We measure eclipse times by first finding a polynomial chain that fits the entire phased light curve, and then we fit the same chain to each successive eclipse, allowing for the temporal shift. The best attained precision of ETVs in {\it Kepler} data is $\sim$6~s \citep{conroy2014}, but this is heavily degraded in case of KIC~8569819 by intrinsic variability that causes variations in eclipse shapes. Nevertheless, variations of a $\sim$76~s peak-to-peak amplitude would be easily detected. Fig.~\ref{fig:etv} depicts the ETV curve and no signal at or around 861~d is detected.

Another technique can independently verify the orbital period of the binary, and evaluate the possibility of a third body in a wider orbit: phase modulation (PM; \citealt{murphyetal14}). The method involves precise determination of the pulsation frequencies of the highest amplitude peaks in the Fourier transform of the stellar light curve, using a non-linear least-squares fit to the Q0--Q17 data with the eclipses masked, and subsequent subdivision of the light curve into smaller segments for analysis. The phase of each peak in each segment is determined, and converted into a light arrival time delay (`time delay', hereafter). The binary motion of the pulsating star should cause an identical signature on each pulsation frequency, with a period equal to the binary orbital period, and an amplitude equal to the light travel time across the projected semi-major axis. Details can be found in \citet{murphyetal14}.

\begin{figure*}
\centering
\includegraphics[width=0.9\linewidth,angle=0]{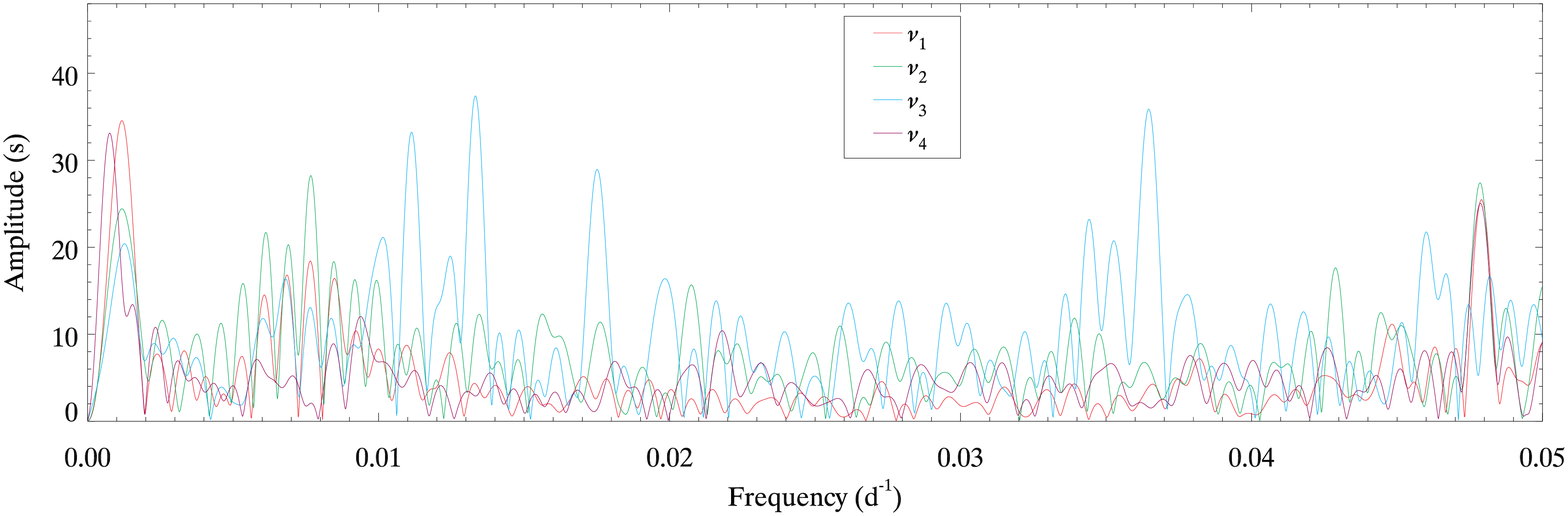}
\includegraphics[width=0.9\linewidth,angle=0]{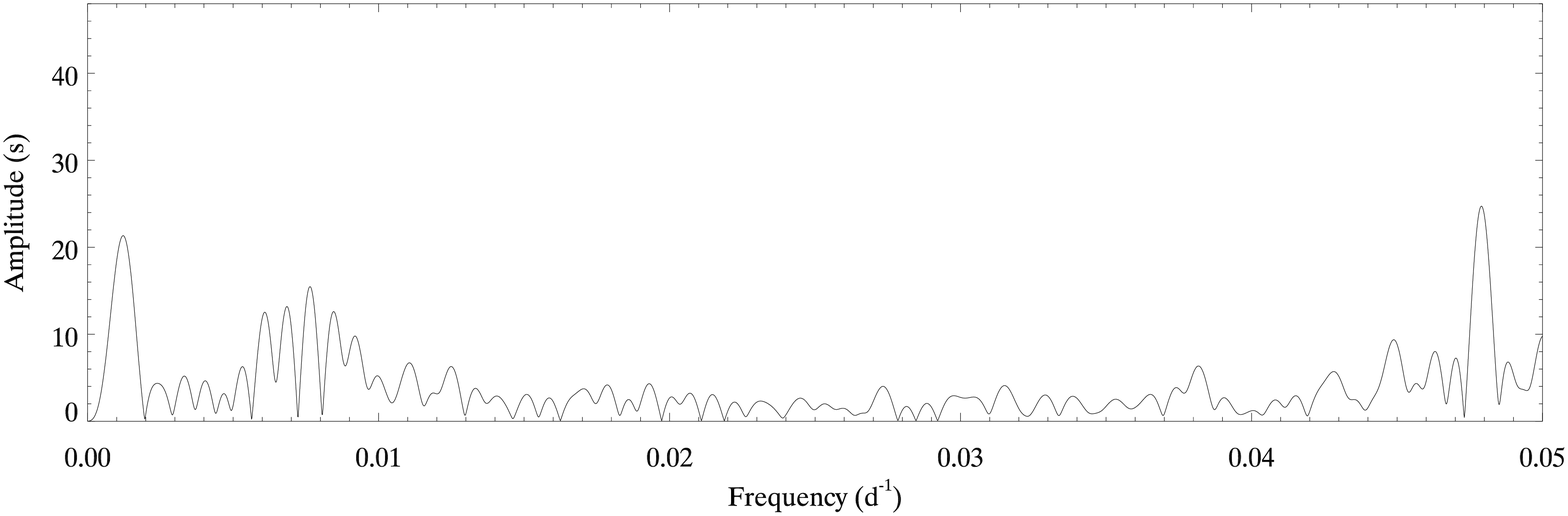}
\caption{Top: the Fourier transform of the light arrival time delays for the four highest-amplitude peaks from Fig.~\ref{fig:8569819_ft-all}. Bottom: the average time delay, weighted by the observed phase uncertainties. The time delays for each of these four peaks agree well on an orbital period of $20.85\pm0.01$~d, as indicated by the peak at 0.048~d$^{-1}$ in the lower panel. However, the lower panel also shows some power at 0.001~d$^{-1}$, originating from the four individual peaks in differing amounts, implicating frequency modulation as the cause. Further discussion is provided in the text.}
\label{fig:time-delays}
\end{figure*}

We chose a segment size of 5~d so that the A star -- G star orbit is well sampled, and we investigated the four highest peaks from Fig.~\ref{fig:8569819_ft-all}. The Fourier transform of the time delays of each peak is shown in the upper panel of Fig.~\ref{fig:time-delays}, where the agreement on the known 20.85-d period of the A star -- G star pair is good. The lower panel of Fig.~\ref{fig:time-delays} is the Fourier transform of the weighted average time delay -- the mean time delay of the four individual peaks, weighted by the phase uncertainties. This shows the orbital frequency of 0.048~d$^{-1}$ (giving $P_{\rm orb}=20.85\pm0.01$~d), but also shows some variability near 0.001~d$^{-1}$. It can be seen in the upper panel that this arises from non-equal contributions from the individual time delays. This illustrates that the long-period variability is not of a binary origin, else each pulsation frequency would respond identically, as in the case for the 20.85-d orbit.

These lines of evidence lead us to conclude that the closely spaced sidelobes to $\nu_1$ represent an intrinsic frequency modulation that is \emph{not} dynamical. The nature of frequency and amplitude modulation in $\delta$~Sct stars is not well understood; this is a topic of current research with the 4-yr {\it Kepler} data sets for thousands of $\delta$~Sct stars \citep{bowman2014}. Importantly, any finding of frequency modulation of pulsation frequencies in pulsating stars must be studied in multiple frequencies for the same star to distinguish between intrinsic and dynamical FM.

We should stress, however, that if the close sidelobes given in Table~\ref{table:lsfit2_8569819_md} and shown in Fig.~\ref{fig:8569819_ft2} had had amplitudes of only 0.020~mmag, which would be a 4$\sigma$ signal, and if they had been dynamical, then the mass function would have given a mass for the third body of about 17~M$_{\rm Jupiter}$. This shows that the FM technique is capable of finding brown dwarfs and gas giant planets in long-period orbits around $\delta$~Sct A stars. Other standard techniques cannot find such objects. A-type stars are too bright for transit detections of small companions; the pulsations mask shallow transits; and such stars are too massive for ground-based radial velocity techniques. The FM technique \citep{FM2012} and the PM technique \citep{murphyetal14} with $\delta$~Sct stars observed by {\it Kepler} have the potential to explore this parameter space for exoplanets.

\section{Binary modelling}
\label{sec:ebmodel}

In this section we present our modelling of the eclipsing binary light curve. This was done independently of the FM analysis presented in the last section, except for the use of the mass function. We do not have a spectroscopic radial velocity curve for this star, hence we use the photometric equivalent of a radial velocity curve, i.e., the FM mass function.

\subsection{Period Analysis}

We performed a period analysis on all the available Q0--17 LC data using the computer package {\sc kephem} \citep{Prsa2011}. {\sc Kephem} is an interactive graphical user interface package that incorporates three methods of period analysis: Lomb-Scargle (LS; Lomb 1976; Scargle 1982),\nocite{Lomb1976, Scargle1982} Analysis of Variance (AoV; Schwarzenberg-Czerny 1989),\nocite{Schwarzenberg-Czerny1989} and Box-fitting Least-Squares (BLS; Kov{\'a}cs et al. 2002),\nocite{Kovacs2002} as implemented in the {\sc vartools} package \citep{Hartmann1998}. Using {\sc kephem} the period and BJD$_0$ (the time of primary minimum) were found, giving an ephemeris:

\begin{equation}
{\rm Min~\Rmnum{1} = BJD2454970.56(1) + 20.84993(3)~d \times E}
\end{equation}

\noindent As can be seen by comparing with the FM analysis in the last section, the light curve fitting of the eclipses gives a more accurate determination of the orbital period.

\subsection{Determination of the eccentricity and argument of periastron through binary star analysis}
\label{sec:param}

To demonstrate the validity of the FM method we generated a binary model to determine the eccentricity and argument of periastron of this system. In our model we assumed the mass of the primary star to be $m_1 = 1.7$~M$_{\odot}$, estimated from the primary star's effective temperature. Consequently, from the mass function determined through FM, we arrived at a mass for the secondary component of $m_2 \sim 1.0$~M$_{\odot}$ (as the system is equator-on). While these assumptions disable the full determination of the binary star parameters, they are adequate to solve robustly for the eccentricity and argument of periastron to validate the FM method.

We applied the binary modelling code {\sc phoebe} \citep{Prsa2005}, which is an extension of the Wilson-Devinney code (\citealt{Wilson1971,Wilson1979}; a manual for the Wilson-Devinney code is available on-line\footnote{ftp://ftp.astro.ufl.edu/pub/wilson/}), to the light curve of KIC~8569819. {\sc phoebe} combines the complete treatment of the Roche potential with the detailed treatment of surface and horizon effects such as limb darkening, reflection and gravity brightening to derive an accurate model of the binary parameters. The current implementation uses the Wilson-Devinney method of summing over the discrete rectangular surface elements, which cover the distorted stellar surfaces, to determine an accurate representation of the total observed flux and consequently a complete set of stellar and orbital parameters. {\sc phoebe} incorporates all the functionality of the Wilson-Devinney code, but also provides an intuitive graphical user interface alongside many other improvements, including updated filters and bindings that enable interfacing between {\sc phoebe} and {\sc python} (see \S~\ref{sec:bayes} below).

The data were detrended using second-order polynomials that were applied between breaks in the {\it Kepler} data, using the {\sc kephem} software. We further cleaned the data by removing all spurious points by eye. To reduce the number of data points, for the purpose of modelling, we assigned each data point with a random number from 0 to 1, and removed all points with random numbers above a specified threshold -- 0.5 during the eclipse phases and 0.01 away from eclipse. This way we retained 50 per cent of the data points during eclipse and 1 per cent of the points away from eclipse. We used a sigmoid function to bridge the number of data points between regions so that discrete changes in the number of data points were avoided. The number of data points was reduced from 60~554 to 4~390. The per-point uncertainty was determined using the standard deviation of the residuals (data minus model) in the out-of-eclipse regions.

Our initial binary model inputs consisted of the effective temperature from the Kepler Input Catalogue (KIC), which we prescribed for the temperature of the primary component ($T_{\rm  eff} = 7100$~K); an estimate of the secondary component's temperature ($T_{\rm  eff} \sim 6100$~K) from consideration of the depths of the eclipses; and the surface gravity value from the KIC $\log g = 4.0$ (cgs units). As the eclipses are separated by $\sim$0.5 in phase, we initially assumed an eccentricity of $e = 0.0$. However, analysis of the relative widths of the eclipses showed that the eccentricity is closer to $e = 0.4$, with an argument of periastron of $\omega \approx 3\upi/2$ implying that we are looking down the line of apsides. 

We assumed pseudo-synchronous rotation, which is stellar rotation synchronous with the orbital velocity at periastron \citep{Hut1981}, and determined the rotation of the components to be $F = 2.715$ rotations per orbit. We assessed the impact of the stellar rotation on the light curve and found an adjustment from $F = 1.0$ to $F = 5.0$ generates a model difference of 0.05~per~cent, which is insignificant. As the Lomb-Scargle method is more accurate than {\sc phoebe} for ephemeris determination, the period and zero point in time were fixed to the values determined using {\sc kephem}. 

When considering the stellar surfaces, we assumed that the primary component has a radiative surface, thus an albedo of $A = 1.0$, and a gravity darkening exponent, $\beta$, of $\beta = 1.0$ \citep{vonZeipel1924}. For the secondary component we assigned the value of $A = 0.6$ for the albedo and $\beta = 0.32$ for the gravity darkening exponent \citep{Lucy1967}. Recent updates in the theory of gravity darkening suggest that this value is dependent on temperature \citep{Claret2011} and/or level of stellar distortion \citep{Espinosa2012}. However, for this system, the gravity darkening value has a negligible effect (0.03~per~cent model difference from $\beta = 0$ to $\beta = 1$, hence the value prescribed by \citet{Lucy1967} was deemed acceptable. Table\,\ref{tab:ParamFix} provides a complete list of the fixed parameters and their assumed values. 

\begin{table} 
\caption{ 
\label{tab:ParamFix} 
\small Fixed parameters and coefficients for the {\sc phoebe} best-fit model to 
the {\it Kepler} light curve for Q0--17. The rotation is specified as a ratio of the stellar rotational to orbital velocity. The mass ratio and semi-major axis were fixed to generate a model with a primary mass of $m_1 = 1.7$~M$_{\odot}$  and a secondary mass of $m_2 = 1.0$~M$_{\odot}$, in line with the mass function determined through the FM method. The fine grid raster is the number of surface elements per quarter of the star at the equator and coarse grid raster is used to determine whether the stars are eclipsing at a given phase.} 
\begin{center} 
\begin{tabular}{||l|r||} 
\hline 
Parameter & Values\\ 
\hline 
Orbital Period (d)                          & 20.84993(3)\\ 
Time of primary minimum BJD$_0$             & 2454970.56(1)\\ 
Primary $T_{\mathrm{eff}}$ (K), $T_1$       & 7100(250)\\
Mass ratio, $q$                             & 0.588\\ 
Semi-major axis (R$_{\odot}$), $a$          & 44.6\\ 
Third light, $l3$                           & 0.0\\ 
Primary rotation, $f1$                      & 2.715\\  
Secondary rotation, $f2$                    & 2.715\\  
Primary Bolometric albedo, $A_1$            & 1.0\\  
Secondary Bolometric albedo, $A_2$          & 0.6\\ 
Primary gravity brightening, $\beta_1$      & 1.0\\ 
Secondary gravity brightening, $\beta_2$    & 0.32\\ 
Primary fine grid raster                    & 90\\  
Secondary fine grid raster                  & 90\\  
Primary coarse grid raster                  & 60\\  
Secondary coarse grid raster                & 60\\  
\hline 
\end{tabular} 
\end{center} 
\end{table}

\begin{table} 
\hfill{} 
\caption{ 
\label{tab:ParamFreePh} 
\small Adjusted parameters and coefficients of the best-fit model to the {\it 
Kepler} light curve for Q0--17. The uncertainties were determined through Markov Chain Monte Carlo methods. The linear and logarithmic limb darkening coefficients are the terms that describe the limb darkening of each component. The limb darkening coefficients were taken from the {\sc phoebe} limb darkening tables \citep{Prsa2011}.}  
\begin{center} 
\begin{tabular}{||l|r||} 
\hline 
Parameter   &{Values}\\ 
\hline 
Phase shift, $\phi$                        & 0.001515(9)\\ 
Orbital eccentricity, $e$                  & 0.366(1)\\ 
Argument of periastron (rad), $\omega$     & 4.72231(8)\\ 
Orbital inclination (degrees), $i$         & 89.91(6)\\ 
T$_{\mathrm{eff}}$ ratio (K), $T_2$/$T_1$  & 0.8517(5)\\
Primary relative luminosity, $L_1$         & 0.873(4)\\ 
Secondary relative luminosity, $L_2$       & 0.1275(6)\\ 
Primary linear limb darkening coefficient       & 0.6169\\ 
Secondary linear limb darkening coefficient     & 0.6382\\ 
Primary logarithmic limb darkening coefficient  & 0.2495\\ 
Secondary logarithmic limb darkening coefficient& 0.2002\\ 
\hline 
\end{tabular} 
\hfill{} 
\end{center} 
\end{table}

\subsubsection{Posterior determination of the orbital parameters \label{sec:bayes}}

\begin{figure*} 
\includegraphics[width=\hsize]{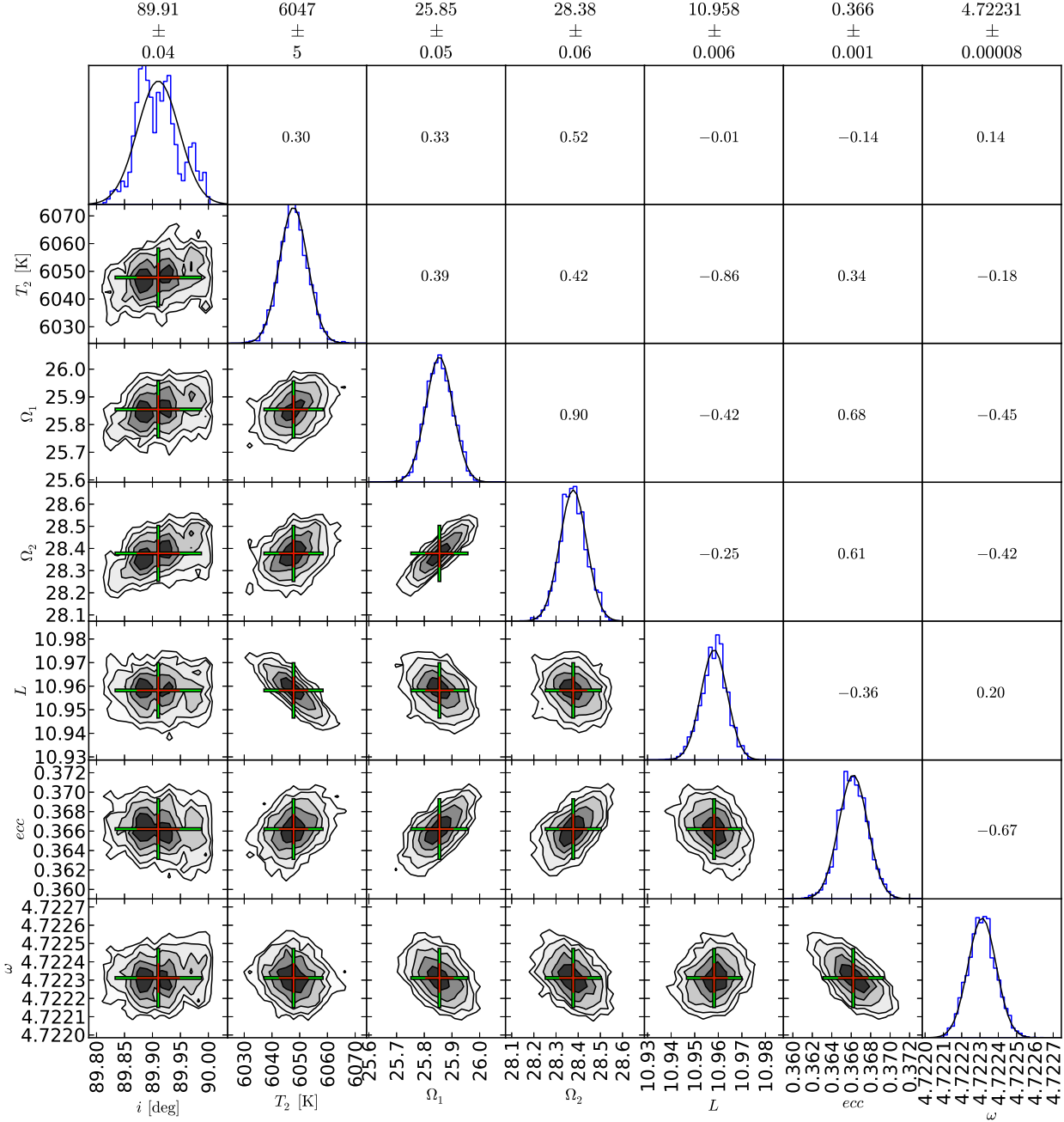}  
\small\caption{Lower left subplots: two dimensional cross-sections of the posterior probability distribution functions. The crosses show the 1$\sigma$ (red) and 2$\sigma$ (green) uncertainties, and are centred on the minima. Diagonal subplots from top left to bottom right: histograms displaying the probability distribution of each individual parameter. Upper right subplots: the correlations for the two-dimensional cross-sections mirrored in the diagonal line where 1 is direct correlation and -1 is a direct anti-correlation. The values above the plot give the mean value and one sigma uncertainty for each parameter, based on the fitted Gaussians.} 
\label{fig:posteriors} 
\end{figure*} 

To determine the posterior probability distribution functions of the binary parameters (cf. Fig.~\ref{fig:posteriors}), we combined {\sc phoebe} with the \mbox{{\sc emcee}}, a {\sc python} implementation of the affine invariant ensemble sampler for Markov chain Monte Carlo (MCMC) proposed by \citet{Goodman2010} and written by \citet{DFM2013}.

\begin{figure*} 
\includegraphics[width=\linewidth]{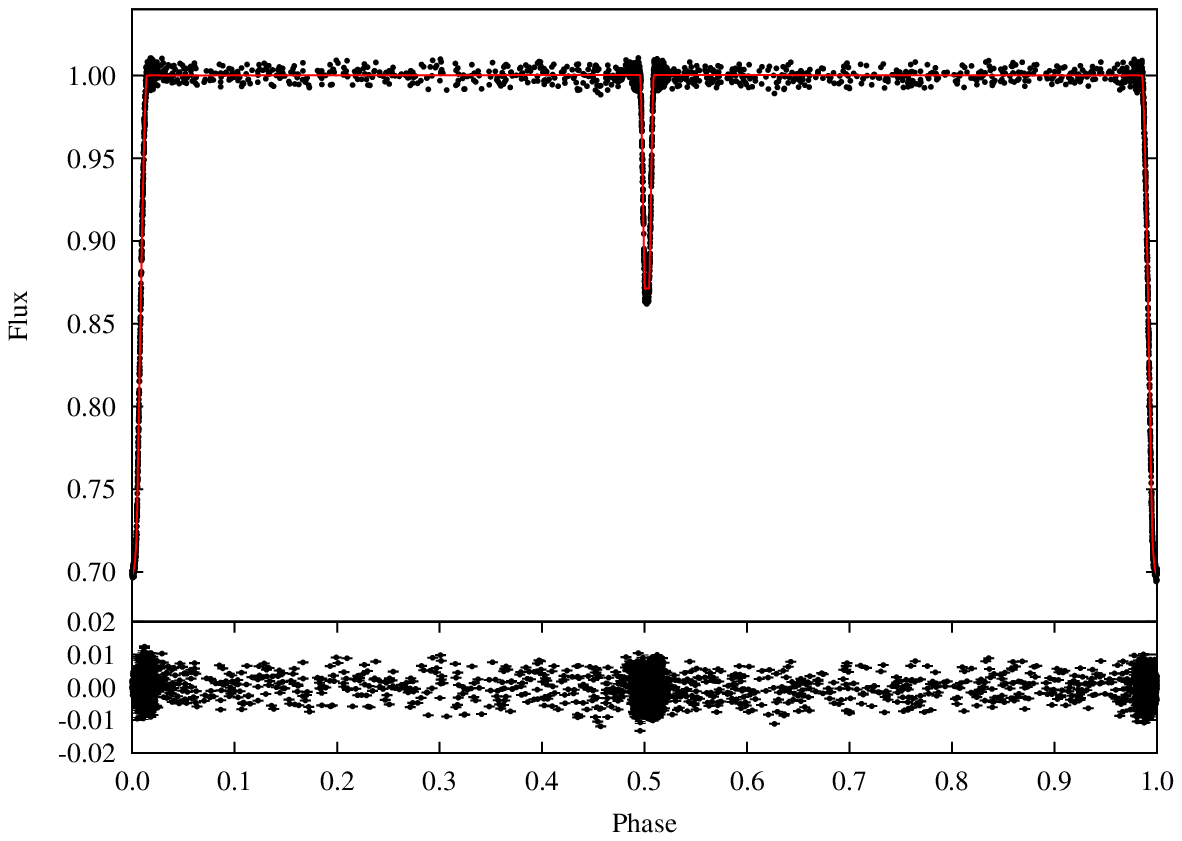}  
\small\caption{Upper panel: theoretical {\sc phoebe} model (red line) and 
observed light curve (black points), prepared as specified in \S\ref{sec:param}. The width of the line depicts the spread of the final 1048 models determined using MCMC. Lower panel: the residuals (black points) of the best-fit model. The per-point uncertainty in the model is displayed as error bars.} 
\label{fig:lc_model} 
\end{figure*} 

MCMC explores the binary parameter space using a set of Markov chains, in this case 128. These chains begin with random distributions based only on the prior probability distribution functions and the initial model. They move through parameter space by assessing their posterior probability distribution function at each point and then selecting a new position based on the position of another chain. The step size is based on the covariance of the two chains. If the move increases the posterior probability distribution function then it is accepted, if the move decreases the probability then it may be accepted (to fully explore the phase space). During the initial burn-in time the Markov chains merge towards their equilibrium position. After this period the chains sample the phase space in terms of their posterior probability distribution functions. The statistics of a large number of iterations ($\sim$150\,000 excluding the burn-in time), provide probability distributions for the model parameters. 

We sampled 7 parameters in our multidimensional parameter space, based on their contribution to the observed flux variation of this system. As only the ratio of the temperatures can be determined from light curve analysis, the effective temperature of the secondary was sampled, whilst keeping the primary temperature fixed. We selected the secondary temperature, since the KIC temperature provides a constraint for the primary effective temperature. The inclination, eccentricity, argument of periastron, primary and secondary potentials (potentials of the Roche lobe -- a proxy for the inverse radius) and luminosity were also sampled using MCMC methods. At each iteration we calculated the phase shift using the new values of eccentricity and argument of periastron. All other parameters in our models were either well determined (period and zero point in time), theoretically determined (albedo and gravity darkening) or insignificant for this system (stellar rotational velocity and gravity darkening). 

For each parameter we used a flat, uniform prior. The prior ranges were selected to be as large as possible without creating unphysical models. We restricted the prior on the inclination to be contained below 90$^{\circ}$ to avoid obtaining a double-peaked distribution reflected about 90$^{\circ}$. The likelihood function was generated by computing the $\chi^2$ difference between the initial model and data. Fig.~\ref{fig:lc_model} shows the average of the last 1024 models generated using MCMC (eight from each Markov chain). The thickness of the line denotes the spread of the last 1024 models. The lower panel shows the residuals to the best fit model. Error bars on the residuals show the per-point standard deviation for the last 1024 models.

The posteriors generated through MCMC are well determined, thus the model is well constrained. For all parameters except the inclination, the Gaussian fit to each posterior, shown in Fig.~\ref{fig:posteriors}, is excellent and provides a robust error estimate. The inclination, however, presents an apparent multimodal distribution, which is a consequence of a small star passing over a large disk. Here the information regarding the points of ingress and egress is limited, yet constrained to a very small range of inclinations. To account for the inexact fit of the Gaussian we have increased the uncertainty of the inclination from that determined through Gaussian fitting, $0.04^{\circ}$ to $0.06 ^{\circ}$.

The stellar potentials and radii are highly dependent on the mass ratio, and thus by assuming the masses we were unable to obtain accurate values. To determine the extent of our assumptions, we perturbed the mass ratio by 10~per~cent and assessed the impact this had on the model. When both increasing and decreasing the mass ratio by 10~per~cent (whilst calculating the potentials to keep the radii fixed) we found a model difference of 0.6~per~cent. As the noise in our data is $\sim$2~per~cent, this difference is not significant. Thus we find that the values reported in Table~\ref{tab:ParamFreePh} are independent of assumption that the mass ratio is $q = 0.588$. 

\section{Conclusions}

We have demonstrated the validity of the frequency modulation (FM) technique \citep{FM2012} by showing the consistent results obtained from it when compared to a traditional eclipsing binary light curve analysis. We derived the mass function from the FM technique of \citet{FM2012}. That additional constraint was then used in the light curve modelling by traditional methods. The orbital period, eccentricity and argument of periastron derived independently from both the FM method and light curve modelling are in good agreement, as is shown in Table~\ref{tab:comp}. This was the primary goal of this paper for readers who are familiar with traditional binary star light curve modelling, but not yet with the FM technique. 

While light curve modelling produces higher accuracy for orbital period and eccentricity in the case of KIC~8569819, that is only true for eclipsing binary stars. For non-eclipsing systems the FM technique is still applicable, whereas traditional techniques work less well for ellipsoidal variables, and not at all for non-distorted, longer orbital period systems. We also have presented a cautionary note in the use of FM in the discovery of additional pure frequency modulation in KIC~8569819 that is not the result of orbital motion, but is intrinsic to the pulsation cavity of the highest amplitude mode in the star. Thus our message is that FM is a powerful technique, but at least two pulsation frequencies in a star must give consistent results to conclude a dynamical origin of the frequency modulation. 

\begin{table}
\centering
\caption[]{Comparison of results from traditional eclipsing binary star light curve modelling and the frequency modulation technique. The agreement validates the FM method for those who are more accustomed to traditional eclipsing binary light curve modelling. }
\begin{tabular}{lcc}
\hline
&\multicolumn{1}{c}{{\sc phoebe}} & \multicolumn{1}{c}{FM}  \\
\hline
orbital period (d) & $20.84993 \pm 0.00003$   &  $20.849 \pm 0.006$\\
eccentricity & $0.366 \pm 0.001$ & $0.39 \pm 0.08$ \\
argument of periastron (rad) & $4.72231 \pm 0.00008$ & $4.59 \pm 0.16$ \\
\hline
\end{tabular}
\label{tab:comp}
\end{table}

\section*{acknowledgements}

\bibliography{arxiv_8569819}

\end{document}